\documentclass{jfm}

\usepackage{graphicx}
\usepackage{epstopdf, epsfig}
\usepackage{newtxtext}
\usepackage{newtxmath}
\usepackage{natbib}
\usepackage{hyperref}
\hypersetup{
    colorlinks = true,
    urlcolor   = blue,
    citecolor  = black,
}

\newcommand{\RomanNumeralCaps}[1]
\linenumbers

\usepackage{upgreek}
\usepackage{amsmath,amssymb,amsfonts,bm}
\usepackage[figuresright]{rotating}
\usepackage{ifpdf}
\usepackage[T1]{fontenc}
\usepackage{xcolor}

\usepackage{epstopdf, epsfig}
\usepackage{color}
\usepackage{tikz}
\usepackage{afterpage}
\usepackage{enumerate}

\definecolor{gray}{rgb}{0.5,0.5,0.5}
\definecolor{c1}{rgb}{0.490,0.796,0.631}
\definecolor{c2}{rgb}{0.271,0.682,0.627}
\definecolor{c3}{rgb}{0.008,0.569,0.600}
\definecolor{c4}{rgb}{0.020,0.443,0.545}
\definecolor{c5}{rgb}{0.020,0.322,0.455}
\definecolor{c6}{rgb}{0.008,0.188,0.278}
\definecolor{c7}{rgb}{0.957,0.647,0.510}
\definecolor{c8}{rgb}{0.572,0.772,0.871}

\usetikzlibrary{shapes}
\usetikzlibrary{plotmarks}

\newcommand{\solidline}[1][black]{\raisebox{2pt}{\tikz{\draw[-,color=#1,solid,line width = 0.5pt](0,0) -- (6mm,0);}}} %make color an argument

\newcommand{\solidlinethick}[1][black]{\raisebox{2pt}{\tikz{\draw[-,color=#1,solid,line width = 2pt](0,0) -- (6mm,0);}}} %make color an argument

\newcommand{\dashedline}[1][black]{\raisebox{2pt}{\tikz{\draw[-,color=#1,dashed,line width = 0.5pt](0,0) -- (6mm,0);}}}

\makeatletter
\usetikzlibrary{calc}
\newcommand*\circled[1][]{\tikz[baseline=(char.base)]{
		\node[shape=circle,draw=black,fill=white,inner sep=0pt,line width=1pt,minimum height={\f@size*1},] (char) {\vphantom{WAH1g}{\footnotesize{\textbf{#1}}}};}}
\makeatother

\graphicspath{{./Figures/}}

\title{Towards decoupling the effects of permeability and roughness on turbulent boundary layers}

\author{D. D. Wangsawijaya\aff{1}
	\corresp{\email{D.D.Wangsawijaya@soton.ac.uk}},
	P. Jaiswal\aff{1,2}
	\and B. Ganapathisubramani \aff{1}}

\affiliation{\aff{1}Aeronautical and Astronautical Engineering, University of Southampton SO17 1BJ, United Kingdom
	\aff{2}Department of Mechanical Engineering, University of Sherbrooke, Sherbrooke, QC, J1K 2R1, Canada}

\begin{document}
	\maketitle
	
	\begin{abstract}
		Boundary layer flow over a realistic porous wall might contain both the effects of wall-permeability and wall-roughness. These two effects are typically examined in the context of a rough-wall flow, i.e., by defining a ``roughness'' length or equivalent to capture the effect of the surface on momentum deficit/drag. In this work, we examine the hypothesis of \cite{esteban2022}, that a turbulent boundary layer over a porous wall could be modelled as a superposition of the roughness effects on the permeability effects by using independently obtained information on permeability and roughness. We carry out wind tunnel experiments at high Reynolds number ($14400 \leq Re_{\tau} \leq 33100$) on various combinations of porous walls where different roughnesses are overlaid over a given permeable wall. Measurements are also conducted on the permeable wall as well as the rough walls independently to obtain the corresponding lengthscales. Analysis of mean flow data across all these measurements suggests that an empirical formulation can be obtained where the momentum deficit ($\Delta U^+$) is modelled as a combination of independently obtained roughness and permeability lengthscales. This formulation assumes the presence of outer-layer similarity across these different surfaces, which is shown to be valid at high Reynolds numbers. Finally, this decoupling approach is equivalent to the area-weighted power-mean of the respective permeability and roughness lengthscales, consistent with the approach recently suggested by \cite{hutchins2023} to capture the effects of heterogeneous rough surfaces.
	\end{abstract}
	
	
	%{\bf MSC Codes }  {\it(Optional)} Please enter your MSC Codes here
	\section{Introduction}
	\label{sec:intro}
	
\begin{figure}
		\centering
		\setlength{\unitlength}{1cm}
		\begin{picture}(8,3)
			\put(0,0){\includegraphics[width=8cm, keepaspectratio]{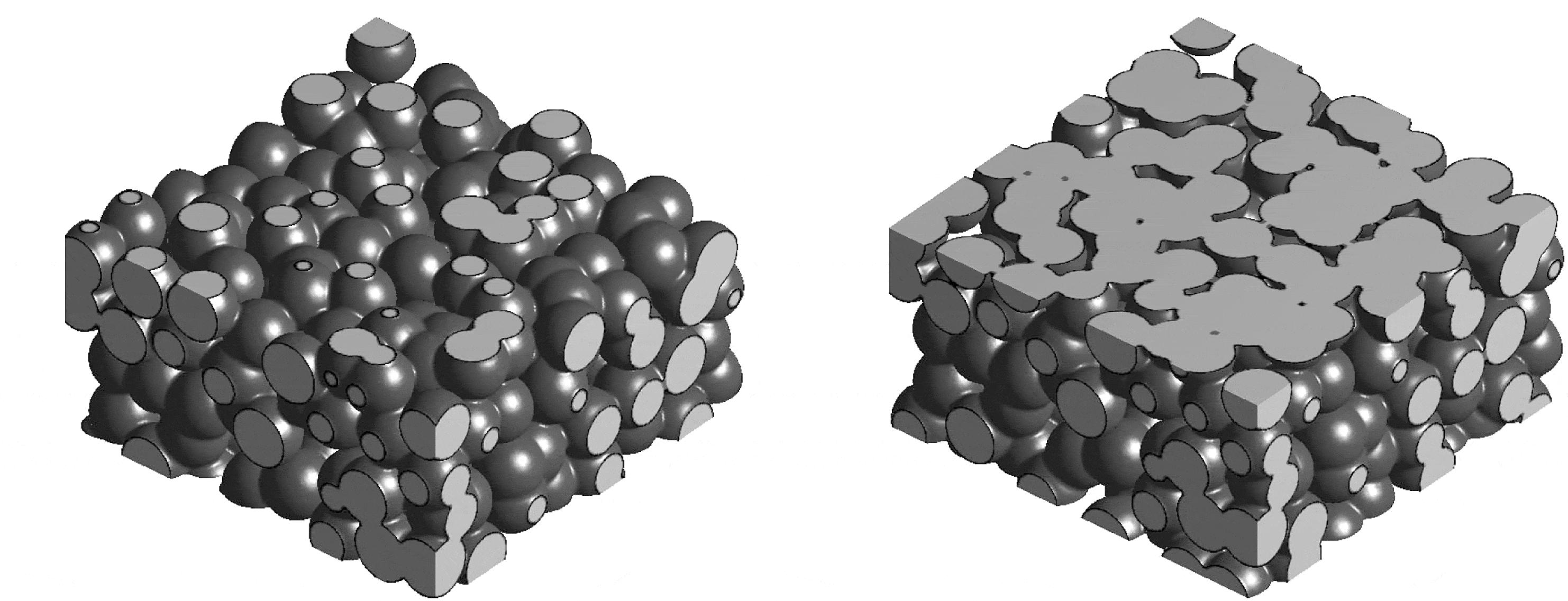}}
			\put(0,3){(a)}
			\put(4,3){(b)}
		\end{picture}
		\caption{Illustrations of realistic permeable walls with: (a) a rough wall on the interface and (b) flat interface. Adapted from \cite{rosti2015}.}
		\label{fig:porous}
	\end{figure}

	Flow over porous walls comprises an extensive number of natural phenomena, ranging from blood vessels to the atmospheric boundary layer (ABL) developing over a forest canopy. The latter can be considered as a turbulent boundary layer (TBL) developing over a porous wall. An example of porous walls, constructed from packed spheres, is shown in figure~\ref{fig:porous}. Here, both walls approximately have the same permeability (resistance of a substrate to fluid flows), but the wall shown in figure~\ref{fig:porous}(a) has a roughness interface above the porous wall, while in figure~\ref{fig:porous}(b) this interface comprises of a flat surface. Thus, a realistic representation of a porous wall is \emph{both} permeable \emph{and} rough to some extent; for the wall in figure~\ref{fig:porous}(a), the effects of both permeability and roughness have to be considered for characterisation of a TBL developing over such wall.
	
	%%%%%%%%%%%%%%%%%%%%%%%%%%%%%%%%%%%%%%%%%%%%%%%%%%%%%%%%%%%%%%%%%%%%%%%%%%%%%%%%%%%%%%%%%%%%%%%%
	\subsection{Rough walls}
	The presence of rough walls increases skin friction from that of smooth walls and thus the time-averaged streamwise velocity $U$ of TBLs developing over rough walls can be written as a downward shift of the logarithmic region $\Updelta U^+$ from the smooth wall velocity profile
	\begin{equation}
		U^+ = \frac{1}{\kappa} \ln \left[ \frac{(y + d) U_{\tau}}{\nu} \right] + B - \Updelta U^+ = \frac{1}{\kappa} \ln (y + y_0)^+ + B - \Updelta U^+
		\label{eq:rough1}
	\end{equation}
	The viscous-scaled velocity is defined as $U^+ \equiv U/U_{\tau}$, where $U_{\tau} \equiv \sqrt{\tau_w/\rho}$ is the friction velocity, $\tau_w$ and $\rho$ are the wall shear stress (WSS) and the density of fluid, respectively. The logarithmic profile is defined as follows: $\kappa$ is the von K{\'a}rm{\'a}n constant, $d$ is the zero-plane displacement, $\nu$ is the kinematic viscosity of fluid, and $B$ is the log-law intercept. The logarithmic shift $\Updelta U^+$, also known as the Hama roughness function \citep{hama1954}, is defined as
	\begin{equation}
		\Updelta U^+ = \frac{1}{\kappa} \ln \left(\frac{k_s U_{\tau}}{\nu}\right) + B - B_{FR} = \frac{1}{\kappa} \ln k_s^+ + B - B_{FR}
		\label{eq:ks}
	\end{equation}
	where $k_s$ is the equivalent sand grain roughness and $B_{FR} = 8.5$ is the `fully' rough intercept of the velocity profile of sand grain roughness. It should be noted that $k_s$ is a measure of roughness effect on the flow relative to that of a uniform sand grain roughness \citep{nikuradse1933}; it can only be determined by flow measurements. The mean streamwise velocity profile of a rough wall can therefore be written as a function of its $k_s$
	\begin{equation}
		U^+ = \frac{1}{\kappa} \ln \left( \frac{y + d}{k_s} \right) + B_{FR}
		\label{eq:rough2}
	\end{equation}
	
	\subsection{Permeable walls}
	
	Earlier studies of TBLs developing over permeable walls involved various types of such walls, namely: packed spheres \citep{zagni1976}, perforated sheets \citep{kong1982}, bed of grains \citep{zippe1983}, and multi-layered walls \citep{manes2009}. Permeable walls have been found to increase drag (i.e. skin friction coefficient $C_f \equiv 2(U_{\tau}/U_{\infty})^2$) from that of solid, impermeable walls, which is attributed to the increase of dissipation, momentum flux, and Reynolds shear stress on the interface between the fluid and the substrates \citep{zagni1976,shimizu1990}. Similar to the rough wall TBLs, the increase in $C_f$ of permeable walls is also characterised by a downward shift in the logarithmic region from that of a solid, smooth wall \citep{hahn2002,efstathiou2018}. Thus, it is probable that there is a ``universal" parameter that characterise permeable walls -- possibly equivalent to $k_s$ for rough wall TBLs. It was suggested by \cite{manes2011} that the logarithmic region of a TBL developing over a permeable wall scales on permeability $K$, and this can be further rearranged to show a downward logarithmic shift,
	\begin{equation}
		U^+ = \frac{1}{\kappa} \ln \left( \frac{y + d}{\sqrt{K}} \right) + c_1 = \frac{1}{\kappa} \ln (y + d)^+ - \frac{1}{\kappa} \ln Re_K + c_1  
		\label{eq:porous1}
	\end{equation} 
	where $Re_K \equiv \sqrt{K} U_{\tau}/\nu$. Previous studies observed a wide range of magnitude of $\kappa$ \citep{breugem2006,manes2011,suga2010}, which is largely attributed to the low $Re$ at which these studies were conducted and thus there was not enough separation between the inner and outer layers of the wall-bounded flows \citep{manes2011}. A more recent experimental work by \cite{esteban2022}, conducted at a higher order of magnitude of $Re$ ($2000 \leq Re_{\tau} \leq 18000$), observed that $\kappa = 0.39$, similar to that of smooth and rough wall TBLs. It was further hypothesised by \cite{esteban2022} that $c_1$ in equation (\ref{eq:porous1}) is an additive constant related to the blockage effect of a porous substrate, as a realistic porous wall comprises of both permeable matrix and solid substrates (see, for example, figure~\ref{fig:porous}), whose size does not permit the full isolation of permeability effect \citep{breugem2006}. This blockage effect might be represented by a roughness function $\Updelta U_b^+ \equiv 1/\kappa \ln k_{s_b}^+ + B - B_{FR}$, similar to that of equation (\ref{eq:ks})  
	\begin{align}
		%U^+ &= \frac{1}{\kappa} \ln (y + d)^+ - \frac{1}{\kappa} \ln Re_K - \Updelta U_b^+ + B \label{eq:porous3}\\
		U^+ &= \frac{1}{\kappa} \ln (y + d)^+ - \frac{1}{\kappa} \ln Re_K - \frac{1}{\kappa} \ln k_{s_b}^+ + B_{FR} = \frac{1}{\kappa} \ln \left( \frac{y + d}{Re_K k_{s_b}} \right) + B_{FR} \label{eq:porous4}
		%U^+ &= \frac{1}{\kappa} \ln \left( \frac{y + d}{Re_K k_{s_b}} \right) + B_{FR} \label{eq:porous5}
	\end{align}   
	where subscript `$b$' denote the blockage effect of the porous substrate. By comparing equation (\ref{eq:rough2}) and (\ref{eq:porous4}), the `equivalent sand grain roughness' $k_s$ of a porous wall (subscript `$p$') can be defined as
	\begin{equation}
		k_{s_p} = Re_K k_{s_b}
		\label{eq:ksp}
	\end{equation}  
	
	In this study, we take the first steps towards exploring the possibility of decoupling permeability and roughness effects on turbulent boundary layers. We have constructed three different porous test surfaces where we maintain the permeability ($Re_K$) but the blockage is systematically altered by adding roughness on to a given permeable substrate. Detailed hot-wire and drag-balance measurements are taken for the permeable, rough and the combination of permeable-rough surfaces over a wide range of Reynolds numbers (ensuring the flow is in fully-rough regime for all cases). We use the experimental data to explore two different approaches for decoupling the two effects. First, we extend the above-mentioned framework to include the effects of additional roughness. Second, we use a power-mean averaging approach proposed for heterogeneous roughness (\citealt{hutchins2023}) to find an equivalent roughness lengthscale that can capture the combined effect of permeability and roughness. 
	
	%surfaces within a range of wide range of $Re_{\tau}$ ($14400 \leq Re_{\tau} \leq 33100$). Hot-wire anemometry (HWA) measures the streamwise velocity component as a function of time and wall-normal location, while the floating element drag balance gives a direct wall shear stress (WSS) measurements. The streamwise velocity of a developing TBL can be decomposed into its time-averaged value ($U$) and the fluctuation about their temporal average ($u'$). The axis system $\bm{x} = (x, y, z)$ corresponds to the streamwise, wall-normal, and spanwise directions, respectively. 

	%%%%%%%%%%%%%%%%%%%%%%%%%%%%%%%%%%%%%%%%%%%%%%%%%%%%%%%%%%%%%%%%%%%%%%%%%%%%%%%%%%%%%%%%%%%%%%%%
	\section{Experimental Setup}
	\label{sec:setup}
	
	%\afterpage{
	\begin{figure}
		\centering
		\setlength{\unitlength}{1cm}
		\begin{picture}(13.5,6)
			\put(9,0){\includegraphics[width=4.5cm, height=6cm, keepaspectratio]{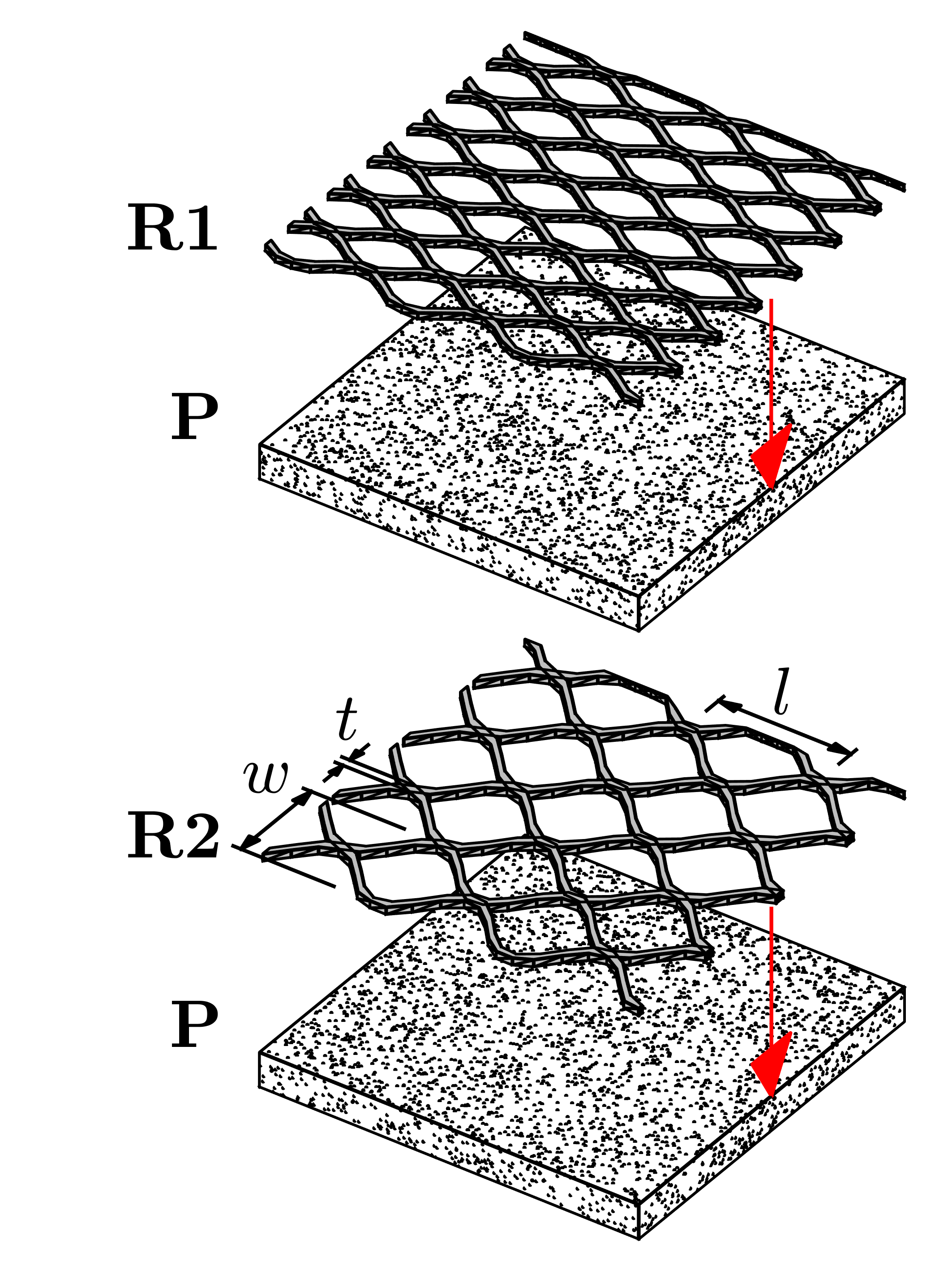}}
			
			\put(0.25,0){\includegraphics[width=9cm, height=6cm, keepaspectratio]{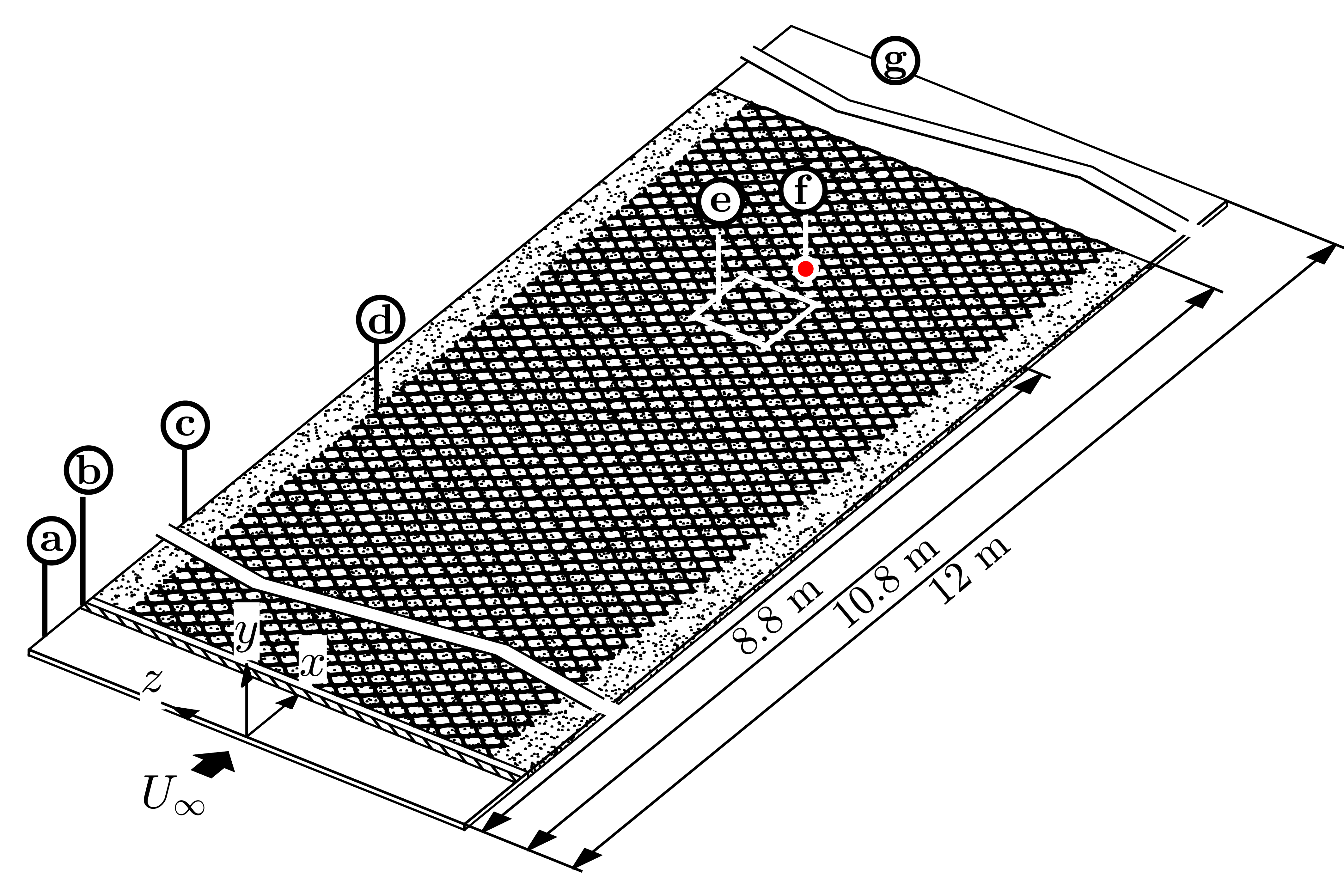}}
			\put(0,3.7){\includegraphics[height=2.3cm, keepaspectratio]{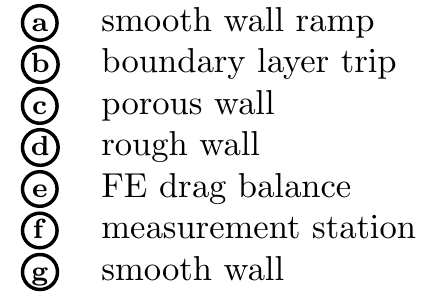}}
			
			\put(4,6){(a)}
			\put(10,6){(b)}
			\put(10,3){(c)}
			\put(1.5,-0.5){(d)}
		\end{picture}
		
		\begin{center}
			\def~{\hphantom{0}}
			\begin{tabular}{c c c c c c c c c}
				%\hline
				& $k$ & $s$ & $\varepsilon$ & $K$ & $A_o$ & $l$ & $w$ & $t$ \\
				Surface & [mm] & [mm] & [--] & [$\times 10^{-8}$ m$^{2}$] & [--] & [mm] & [mm] & [mm] \\%[3pt]
				\hline
				
				\textbf{P} & 15 & 0.89 & 0.859 & 3.62 & -- & -- & -- & -- \\
				\textbf{R1} & 3.0 & -- & -- & -- & 0.73 & 62 & 30 & 4 \\ 
				\textbf{R2} & 3.5 & -- & -- & -- & 0.81 & 70 & 53 & 5 \\ 
				%\hline
			\end{tabular}
			\caption{(a) Illustration of a test surface laid inside the test section of the boundary layer wind tunnel. Combination of porous-rough test surfaces: (b) porous \textbf{P}--rough wall \textbf{R1}, and (c) porous \textbf{P}--rough wall \textbf{R2}. (d) Geometric parameters of \textbf{P}, \textbf{R1}, and \textbf{R2}, where $k$ is the thickness of the surfaces. For \textbf{P}: $s$ is the average pore size of the foam, $\varepsilon$ is the porosity, and $K$ is the permeability, obtained from \cite{esteban2022}. For \textbf{R1} and \textbf{R2}: $A_o$ is the open area ratio of the mesh, $l$ is the length of the longway of the mesh, $w$ is the length of the shortway of the mesh, and $t$ is the width of the mesh strand, as illustrated in (c).}
			\label{fig:setup}
		\end{center}
	\end{figure}
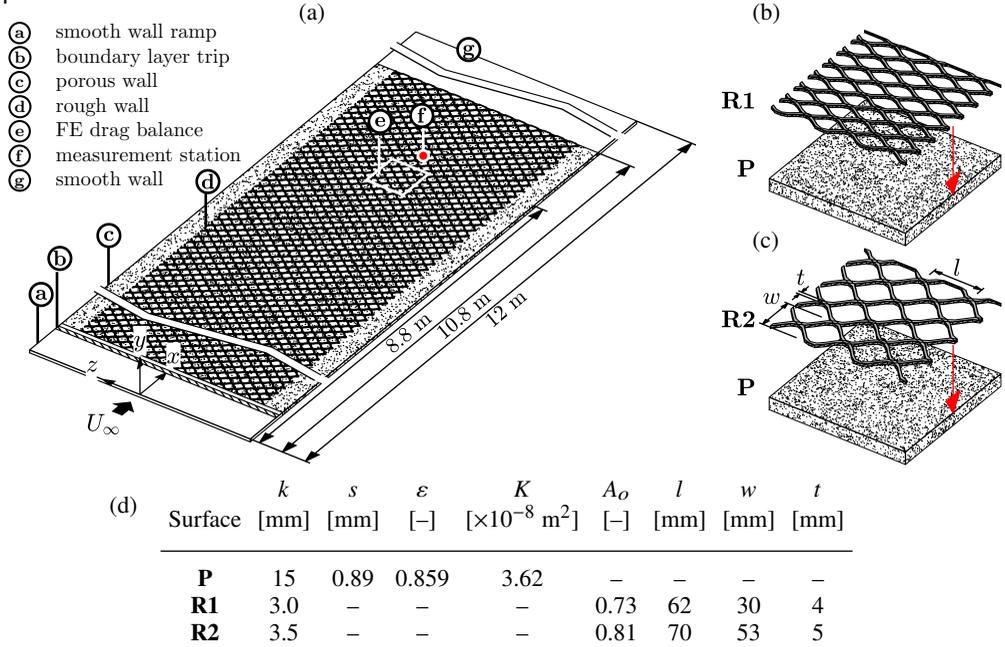
	%}
	
	%%%%%%%%%%%%%%%%%%%%%%%%%%%%%%%%%%%%%%%%%%%%%%%%%%%%%%%%%%%%%%%%%%%%%%%%%%%%%%%%%%%%%%%%%%%%%%%%
	\subsection{Facility}
	\label{sub:facility}
	
	Measurements are conducted inside the closed return boundary layer wind tunnel (BLWT) at the University of Southampton. The flow passes through a contraction section of 6:1 ratio before entering the 12 m $\times$ 1.2 m $\times$ 1 m (length $\times$ width $\times$ height) test section. The boundary layer develops over the floor (bottom surface) of the BLWT. A 220 mm-long smooth wall ramp is installed at the end of the contraction section (figure~\ref{fig:setup}(a)\circled[a]) to match the thickness of the test surfaces. This ramp marks the inlet towards the test section of the BLWT and the streamwise datum ($x = 0$). The boundary layer is tripped by a 8.5 mm-wide, 0.4 mm-thick turbulator tape attached on the ramp (figure~\ref{fig:setup}(a)\circled[b]). The tunnel has the freestream turbulence level of $\sigma_{u'}/U_{\infty} \approx 0.1$\%, where $\sigma_{u'}$ is the standard deviation of streamwise turbulent fluctuation at the freestream and $U_{\infty}$ is the freestream velocity. The BLWT is equipped with a water-cooled heat exchanger, maintaining the flow temperature variation of 1\% for the longest measurements ($\sim$8 hours).   
	
	%The roof (top surface) of the BLWT is non-adjustable. Thus, the measurements for all test surfaces are conducted in nominally favourable pressure gradients, as indicated by the magnitude of the wake parameter $\Pi$ of the wake function of the mean velocity profile of \cite{coles1956}, $2\Pi/\kappa W(y/\delta)$. For all test surfaces, $\Pi = 0.18 \pm 0.16$ for all test surfaces (table~\ref{tab:case}), which is less than that of zero pressure gradient TBLs, $\Pi \approx 0.4$ \citep{chauhan2009,marusic2015}. $\Pi$ is estimated by fitting the obtained velocity profile to the composite profile of \cite{rodriguezlopez2015}.
	
	%%%%%%%%%%%%%%%%%%%%%%%%%%%%%%%%%%%%%%%%%%%%%%%%%%%%%%%%%%%%%%%%%%%%%%%%%%%%%%%%%%%%%%%%%%%%%%%%
	\subsection{Test surfaces}  
	
	\begin{table}
		\begin{center}
			\def~{\hphantom{0}}
			\begin{tabular}{l c c c c c c c c c c c c c c}
				Case & $\delta$  & $\theta$ & $U_{\infty}$ & $U_{\tau}$ & $Re_x$ & $Re_{\theta}$ & $Re_{\tau}$ & $Re_K$ & $C_f$ & $\Pi$ & $\Updelta U^+$ & $k_s$ & $k_s^+$ & Sym. \\
				& \scriptsize{[mm]} & \scriptsize{[mm]} & \scriptsize{[ms$^{-1}$]} & \scriptsize{[ms$^{-1}$]} & \scriptsize{[$\times 10^7$]} & \scriptsize{[$\times 10^4$]} & \scriptsize{[$\times 10^4$]} & \scriptsize{[--]} & \scriptsize{[$\times 10^{-3}$]} & \scriptsize{[--]} & \scriptsize{[--]} & \scriptsize{[mm]} & \scriptsize{[--]} &  \\
				\hline 
				
				%Smooth wall========================================================================
				%\multicolumn{15}{c}{\textbf{Smooth walls}}\\[3pt]
				
				%\textbf{S~~} & 122.21 & 11.96 & 39.51 & 1.30 & 2.32 & 3.15 & 1.06 & -- & 2.18 & 0.32 & -- & -- & -- & \includegraphics[height=2mm, keepaspectratio]{mark_sq1} \\[3pt]
				%\textbf{S~~} & 128.63 & 11.95 & 44.66 & 1.46 & 2.62 & 3.55 & 1.25 & 2.14 & 0.31 & -- & -- & -- & \textcolor{c2}{$\bm \square$}\\[3pt]
				
				%Porous wall========================================================================
				\multicolumn{15}{c}{\textbf{Porous walls}}\\%[3pt]
				
				& 201.11 & 24.55 & 19.84 & 1.06 & 1.18 & 3.29 & 1.44 & 13.62 & 5.72 & 0.27 & 12.01 &  & 560 & \includegraphics[height=2mm, keepaspectratio]{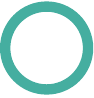}\\
				& 199.96 & 24.45 & 25.45 & 1.36 & 1.52 & 4.21 & 1.85 & 17.57 & 5.74 & 0.27 & 12.66 &  & 723 & \includegraphics[height=2mm, keepaspectratio]{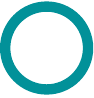}\\
				\textbf{P} & 208.26 & 25.37 & 29.06 & 1.54 & 1.76 & 5.07 & 2.21 & 20.22 & 5.65 & 0.31 & 13.02 & 7.83 & 832 & \includegraphics[height=2mm, keepaspectratio]{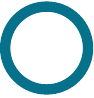}\\
				& 211.27 & 25.88 & 35.46 & 1.87 & 2.14 & 6.30 & 2.72 & 24.50 & 5.59 & 0.34 & 13.52 &  & 1008 & \includegraphics[height=2mm, keepaspectratio]{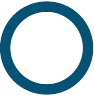}\\
				& 221.25 & 26.43 & 40.05 & 2.13 & 2.39 & 7.18 & 3.20 & 27.52 & 5.67 & 0.24 & 13.81 &  & 1132 & \includegraphics[height=2mm, keepaspectratio]{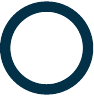}\\ [3pt]
				
				%Rough wall=========================================================================
				\multicolumn{15}{c}{\textbf{Rough walls}}\\%[3pt]
				
				& 232.40 & 28.51 & 14.90 & 0.86 & 0.87 & 2.82 & 1.33 & -- & 6.69 & 0.19 & 12.88 & & 773 & \includegraphics[height=2mm, keepaspectratio]{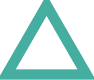} \\
				& 236.84 & 28.78 & 19.93 & 1.15 & 1.17 & 3.83 & 1.82 & -- & 6.70 & 0.18 & 13.64 & & 1041 & \includegraphics[height=2mm, keepaspectratio]{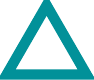} \\
				\textbf{R1} & 239.38 & 28.70 & 25.20 & 1.46 & 1.48 & 4.82 & 2.32 & -- & 6.67 & 0.19 & 14.23 & 13.52 & 1310 & \includegraphics[height=2mm, keepaspectratio]{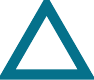} \\
				& 243.78 & 28.84 & 29.58 & 1.71 & 1.73 & 5.68 & 2.77 & -- & 6.69 & 0.18 & 14.65 & & 1539 & \includegraphics[height=2mm, keepaspectratio]{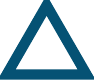} \\
				& 254.33 & 29.52 & 35.21 & 2.04 & 2.05 & 6.87 & 3.42 & -- & 6.69 & 0.15 & 15.08 & & 1821 & \includegraphics[height=2mm, keepaspectratio]{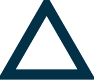} \\%[3pt]
				
				& 241.11 & 29.41 & 15.10 & 0.89 & 0.89 & 2.96 & 1.44 & -- & 7.01 & 0.19 & 13.44 & & 963 & \includegraphics[height=3mm, keepaspectratio]{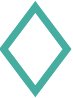} \\
				\textbf{R2} & 229.33 & 28.56 & 19.74 & 1.18 & 1.15 & 3.75 & 1.79 & -- & 7.09 & 0.18 & 14.14 & 16.17 & 1263 & \includegraphics[height=3mm, keepaspectratio]{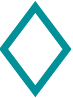} \\
				& 229.34 & 28.25 & 24.90 & 1.49 & 1.46 & 4.68 & 2.27 & -- & 7.12 & 0.18 & 14.74 & & 1599 & \includegraphics[height=3mm, keepaspectratio]{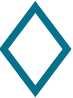} \\
				& 230.03 & 28.64 & 29.83 & 1.78 & 1.75 & 5.69 & 2.73 & -- & 7.13 & 0.16 & 15.21 & & 1920 & \includegraphics[height=3mm, keepaspectratio]{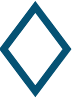} \\ [3pt]
				
				%Porous-rough wall==================================================================
				\multicolumn{15}{c}{\textbf{Porous-rough walls}}\\%[3pt]
				
				& 225.41 & 26.90 & 17.28 & 1.06 & 1.01 & 3.08 & 1.58 & 13.32 & 7.48 & 0.18 & 14.08 & & 1235 & \includegraphics[height=3mm, keepaspectratio]{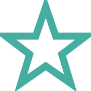}\\ 
				& 218.12 & 27.17 & 21.47 & 1.31 & 1.26 & 3.90 & 1.92 & 16.74 & 7.49 & 0.15 & 14.66 & & 1552 & \includegraphics[height=3mm, keepaspectratio]{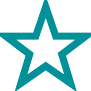}\\ 
				\textbf{PR1} & 227.93 & 27.28 & 26.04 & 1.60 & 1.54 & 4.78 & 2.46 & 20.55 & 7.60 & 0.14 & 15.19 & 17.65 & 1906 & \includegraphics[height=3mm, keepaspectratio]{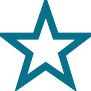}\\ 
				& 226.64 & 27.33 & 31.49 & 1.94 & 1.85 & 5.74 & 2.94 & 24.65 & 7.60 & 0.09 & 15.65 & & 2286 & \includegraphics[height=3mm, keepaspectratio]{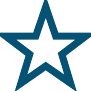}\\ 
				& 229.11 & 27.28 & 35.01 & 2.17 & 2.05 & 6.36 & 3.31 & 27.46 & 7.67 & 0.10 & 15.93 & & 2547 & \includegraphics[height=3mm, keepaspectratio]{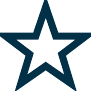}\\ 
				%[3pt]
				
				& 219.42 & 27.29 & 15.06 & 0.95 & 0.90 & 2.79 & 1.41 & 12.27 & 7.94 & 0.13 & 14.23 & & 1254 & \includegraphics[height=3mm, keepaspectratio]{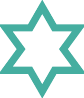}\\ 
				
				\textbf{PR2} 
				& 221.73 & 27.41 & 19.97 & 1.26 & 1.20 & 3.75 & 1.91 & 16.40 & 7.93 & 0.13 & 14.98 & 19.45 & 1677 & \includegraphics[height=3mm, keepaspectratio]{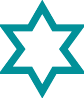}\\ 
				
				& 224.41 & 27.61 & 25.20 & 1.59 & 1.52 & 4.77 & 2.45 & 20.76 & 7.98 & 0.13 & 15.58 & & 2122 & \includegraphics[height=3mm, keepaspectratio]{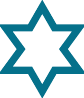}\\ 
				
				& 222.96 & 28.06 & 28.26 & 1.79 & 1.69 & 5.40 & 2.71 & 23.16 & 8.00 & 0.14 & 15.86 & & 2368 & \includegraphics[height=3mm, keepaspectratio]{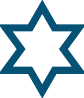}\\ 
				
			\end{tabular}
			\caption{Summary of all test surfaces: baseline smooth walls (\textbf{S}), porous (\textbf{P}), rough (\textbf{R1} and \textbf{R2}), and porous-rough (\textbf{PR1} and \textbf{PR2}) walls. The statistics are obtained from HWA measurements: $\delta$ is the 99\% boundary-layer thickness and $\theta$ is the momentum thickness. Reynolds number definitions are: $Re_{x} \equiv xU_{\infty}/\nu$, $Re_{\theta} \equiv \theta U_{\infty}/\nu$, $Re_{\tau} \equiv \delta U_{\tau}/\nu$, and $Re_{K} \equiv \sqrt{K}U_{\tau}/\nu$. Coefficient of friction is defined as $C_f \equiv 2 (U_{\tau}/U_{\infty})^2$. The last column shows the symbol associated with each test surface. Colours denote test cases with approximately matched $Re_{\tau}$ and $Re_K \approx$ 14400 and 13 (\includegraphics[height=2mm, keepaspectratio]{mark_circ2-eps-converted-to.pdf}), 18600 and 17 (\includegraphics[height=2mm, keepaspectratio]{mark_circ3-eps-converted-to.pdf}), 23400 and 20.5 (\includegraphics[height=2mm, keepaspectratio]{mark_circ4-eps-converted-to.pdf}), 27700 and 24 (\includegraphics[height=2mm, keepaspectratio]{mark_circ5-eps-converted-to.pdf}), 33100 and 27.5 (\includegraphics[height=2mm, keepaspectratio]{mark_circ6-eps-converted-to.pdf}). }
			\label{tab:case}
		\end{center}
	\end{table} 
	
	Five test surfaces are constructed for this study: porous wall (denoted by `\textbf{P}', figure~\ref{fig:setup}(a)\circled[c]), two types of rough walls (`\textbf{R1}' and `\textbf{R2}', figure~\ref{fig:setup}b,c), and two combinations of rough walls on top of the porous wall (`\textbf{PR1}' and `\textbf{PR2}', figure~\ref{fig:setup}b,c). All test surfaces are assembled on the bottom surface of the BLWT, downstream of the ramp and trip (figure~\ref{fig:setup}a). The ramp and test surfaces only cover the first 10.8 m-long part of the test section, while the section from $x = 10.8$ m to $x = 12$ m comprises of a smooth wall (figure~\ref{fig:setup}a). All measurements are conducted at a constant $x$ location at a range freestream velocity $U_{\infty}$, corresponding to a set of matched friction Reynolds number $Re_{\tau} \equiv \delta U_{\tau}/\nu \pm 1600$ (here, $\delta$ is the 99\% boundary-layer thickness) between \textbf{P}, \textbf{R1}, and \textbf{PR1}, as well as another set of matched $Re_{\tau}$ for \textbf{P}, \textbf{R2}, and \textbf{PR2}. The first set of test surfaces has 5 matched $Re_{\tau}$ cases ($Re_{\tau} \approx 14400$, 18600, 23400, 22900, and, 33100), while the second set has four matched cases ($Re_{\tau} \approx 14400$, 18600, 23400, and 27700). Details of each test surfaces, including the statistics obtained from HWA and WSS measurements, are given in table~\ref{tab:case}. Throughout this study, each test surface is denoted with a symbol (see the last column in table~\ref{tab:case}). The same colour denotes surfaces with matched $Re_{\tau}$ and $Re_K$ (within $Re_K \pm 10\%$), which runs from lighter towards darker colour as $Re_{\tau}$ and $Re_K$ increases.   
	
	The porous walls are constructed from sheets of 15 mm-thick ($1060 \leq k^+ \equiv k U_{\tau}/\nu \leq 2130$), 45 ppi (pores per inch) polyurethane reticulated foam. The substrate has the average pore size of $s = 0.89$ mm ($63 \leq s^+ \equiv s U_{\tau}/\nu \leq 126$) and permeability of $K = 3.62 \times 10^{-8}$ m$^{2}$, corresponding to $13 \leq Re_K \leq 27.5$. The rough walls are constructed from two types of diamond-shaped, expanded metal mesh sheets denoted by \textbf{R1} and \textbf{R2} (figure~\ref{fig:setup}b,c). The mesh sheets have the thickness of $k = 3$ and 3.5 mm ($178 \leq k^+ \leq 476$) for \textbf{R1} and \textbf{R2}, respectively, and the open area $A_o$ (ratio of empty to total area in $xz$--plane) of 0.73 and 0.81. Both mesh sheets only cover a 1 m-wide portion (in $z$) of the working section (figure~\ref{fig:setup}(a)\circled[d]). Details of the relevant parameters regarding the porous substrate and mesh geometries are given in figure~\ref{fig:setup}(d).  
	
	%%%%%%%%%%%%%%%%%%%%%%%%%%%%%%%%%%%%%%%%%%%%%%%%%%%%%%%%%%%%%%%%%%%%%%%%%%%%%%%%%%%%%%%%%%%%%%%%
	\subsection{Hot-wire anemometry}
	
	Hot-wire anemometry (HWA) measurements are conducted for all test surfaces listed in table~\ref{tab:case} by traversing the wire in wall-normal direction $y$ across 35--40 logarithmically-spaced points from the wall towards the freestream. All measurements are conducted at the centreline of the tunnel at $x = 8.8$ m from the datum of the tunnel test section (figure~\ref{fig:setup}(a)\circled[f]). A modified Dantec 55P05 single-sensor with a boundary-layer type probe (and an appropriate probe support) are used with a Streamline Pro Constant Temperature Anemometer (CTA) with an overheat ratio of 0.8. The sensor is a 5 $\upmu$m diameter, 1-mm long tungsten wire soldered to the tip of the hot-wire prong, satisfying the recommended wire length-to-diameter ratio of 200 \citep{ligrani1987} and corresponding to the viscous-scaled sensor lengths of $57 \leq l_w^+ \equiv l_w U_{\tau}/\nu \leq 145$. The output signal is sampled from the CTA at $f_s= 30$ kHz, yielding viscous-scaled sampling interval of $1.6 \leq t^+ \equiv  U_{\tau}^2/(f_s \nu) \leq 10$. The sampling time $T_s$ at each measurement point differs between test surfaces such that the boundary-layer turnover time is maintained at $T_s U_{\infty}/\delta \geq 20000$ to allow convergence of turbulence energy spectra \citep{hutchins2009}. The sensor is traversed to the freestream and calibrated prior to and after each measurement. The relation between freestream velocity $U_{\infty}$ and output voltage of the sensor $E_{\infty}$ is defined by King's law $E_{\infty}^2 = C_1 + C_2 U_{\infty}^{C_3}$. Temperature compensation is applied to the output signal to account for a slight sensor drift. 
	
	%%%%%%%%%%%%%%%%%%%%%%%%%%%%%%%%%%%%%%%%%%%%%%%%%%%%%%%%%%%%%%%%%%%%%%%%%%%%%%%%%%%%%%%%%%%%%%%
	\subsection{Wall shear stress (WSS) measurements}
	\label{sub:wss}
	
	\begin{figure}
		\centering
		\setlength{\unitlength}{1cm}
		\begin{picture}(13.5,9.5)
			\put(0,0){\includegraphics[width=13.5cm, keepaspectratio]{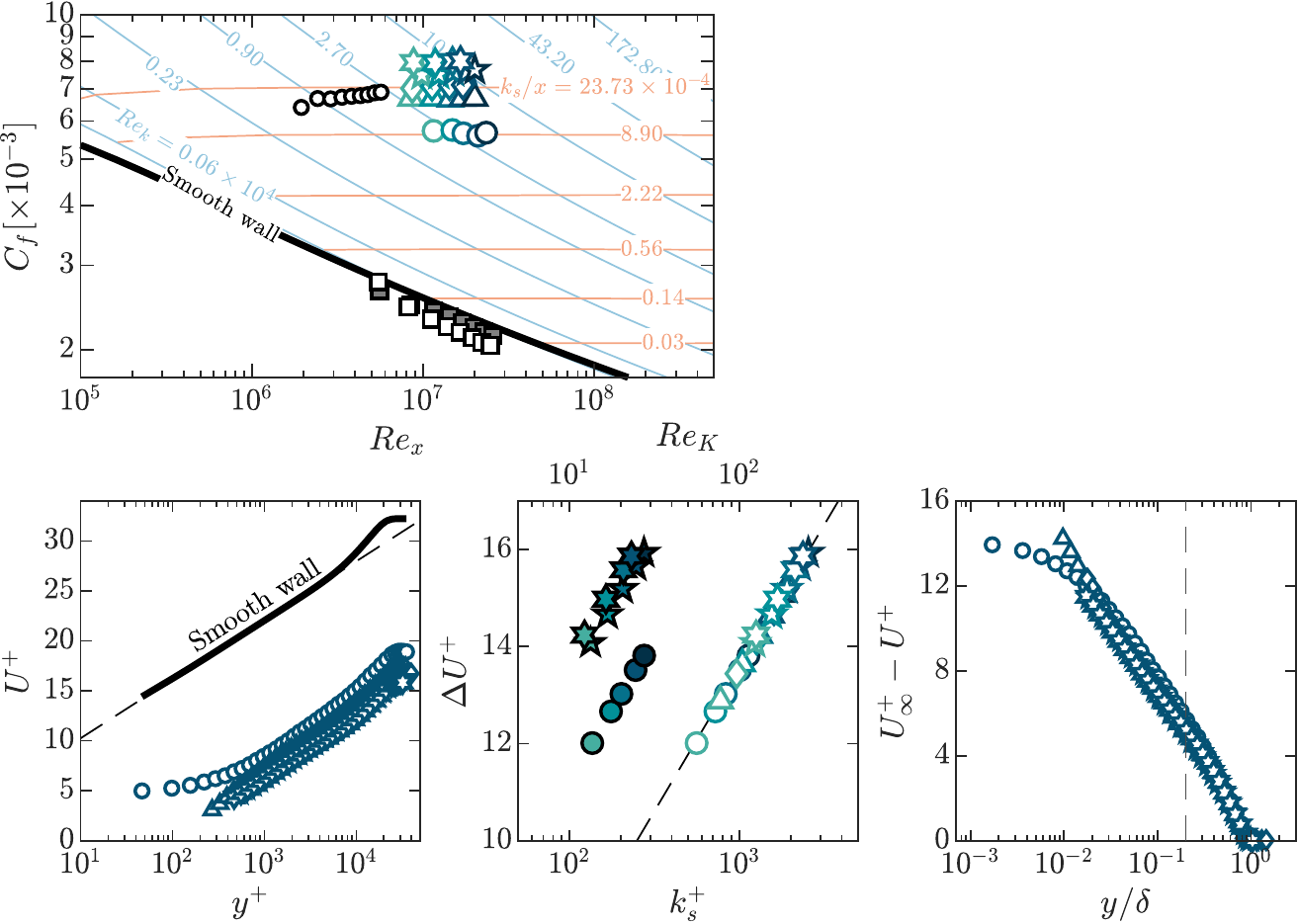}}
			
			\put(9,8){\begin{tabular}{c c c}
					Parameter &  \textbf{PR1} &\textbf{PR2}\\
					\hline
					$k_{s_{pr}}/k_{s_p}$ & 2.25 & 2.48\\
					$C$ & 0.17 & 0.15\\
					$n$ & 1.42 & 1.54
			\end{tabular}}
			
			\put(0,9.5){(a)}
			\put(0,4.5){(b)}
			\put(4.75,4.5){(c)}
			\put(9,4.5){(d)}
			\put(8.5,9){(e)}
			
		\end{picture}
		
		\caption{(a) $C_f$ as a function of $Re_x$ for all test surfaces. Validation of the WSS measurements with smooth walls (\includegraphics[height=2mm, keepaspectratio]{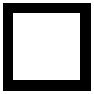}): fitted to the composite profile of \cite{rodriguezlopez2015} (\includegraphics[height=2mm, keepaspectratio]{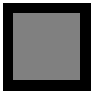}) and the analytical solution of \cite{monty2016} (\solidlinethick). The same porous surface measured by \cite{esteban2022} (\includegraphics[height=2mm, keepaspectratio]{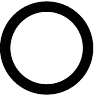}). Porous wall ($k_s = 7.83$ mm) at constant $Re_k \equiv k U_{\infty}/\nu$ (\solidline[c8]) and constant $k_s/x$ (\solidline[c7]), obtained from \cite{monty2016}. (b) $U^+$ as a function of $y^+$ for matched $Re_{\tau} \approx 27700$, \dashedline: $1/\kappa \ln y^+ +B$. (c) $\Updelta U^+$ as functions of $k_s^+$ (white-filled symbols, bottom axis) and $Re_K$ (filled symbols, top axis). \dashedline: $1/\kappa \ln k_s^+ + B - B_{FR}$. Legends are shown in table~\ref{tab:case}. (d) Velocity defect $U_{\infty}^+ - U^+$ as a function of $y/\delta$ for matched $Re_{\tau} \approx 27700$. (e) Decoupling of permeability and roughness effect using methods described in \S\ref{sub:decouple}.}
		\label{fig:Cf}
	\end{figure}
	
	The WSSs of all test surfaces shown in table~\ref{tab:case} are measured using an in-house floating element (FE) drag balance. The balance has a 200 mm $\times$ 200 mm FE located at $x = 8.6$ m (figure~\ref{fig:setup}(a)\circled[e]), slightly upstream of the HWA measurement station. A section of each test surfaces are cut according to the size of the FE and mounted on top of the FE, leaving a 0.5 mm clearance between the FE and its housing, allowing it to move freely for WSS measurements. The housing is sealed to prevent airflow into the balance. Measurements are acquired at $f_s = 256$ Hz with 120 s sampling time for each test surface. Calibration of the balance is performed before measurements by loading a set of calibration weights to the balance via a pulley system. The relationship between known calibration weights and time-averaged output voltage (within $\pm 0.96\%$ uncertainties) from the balance is obtained by fitting the calibration data into a first order polynomial. 
	
	%Due to the nominally favourable pressure gradient of the developing TBL in the BLWT (see \S\ref{sub:facility}), the measured force $mg$ is corrected for non-zero pressure gradient $\mathrm{d}p/\mathrm{d}x$ measured by two static pressure taps attached upstream and downstream of the FE, respectively. 
	
	Figure~\ref{fig:Cf}(a) shows the $C_f$ of all tests surfaces as a function of fetch Reynolds number $Re_x \equiv x U_{\infty}/\nu$. For validation, $C_f$ of smooth walls ($2930 \leq Re_{\tau} \leq 12500$) are determined first using three different methods: (i) direct WSS measurements using the FE drag balance (\includegraphics[height=2mm, keepaspectratio]{mark_sq7-eps-converted-to.pdf}), (ii) fitting velocity profiles obtained from HWA measurements to the composite profile of \cite{rodriguezlopez2015} (\includegraphics[height=2mm, keepaspectratio]{mark_sq8-eps-converted-to.pdf}), (iii) the analytical solution of \cite{monty2016} (\solidlinethick). All three methods show a reasonable agreement with each other to within 5\%.
	
	For the rest of the test surfaces \textbf{P}, \textbf{R}, and \textbf{PR}, $C_f$ are determined from direct WSS measurements. Both rough walls (\textbf{R1} and \textbf{R2}) have higher $C_f$ than that of porous walls, which increase further as the walls are combined with the porous walls (\textbf{PR1} and \textbf{PR2}), as shown in figure~\ref{fig:Cf}(a). The logarithmic shift $\Updelta U^+$ and $k_s$ for these surfaces are obtained by fitting the mean profile to a modified method of \cite{rodriguezlopez2015}, in which the log-law shift term $\Updelta U^+$ is added, forming equation (\ref{eq:rough1}) while maintaining the $U_{\tau}$ obtained from direct WSS measurements. The fit yields $\kappa = 0.39$ and $B = 4.34 \pm 0.08$ (consistent with those obtained for TBLs developing over various porous walls measured by \citealp{esteban2022}). Porous wall $C_f$ (figure~\ref{fig:Cf}a) is compared to the analytical solution proposed by \cite{monty2016} over constant $k_s/x = 8.9 \times 10^{-4}$ ($k_s = 7.83$ mm). Additionally, $C_f$ obtained by \cite{esteban2022} for the same porous wall, $k_s/x = 2.4 \times 10^{-3}$ ($k_s = 7.97$ mm) is also compared with the analytical solution, both showing good agreement with these solutions. 
	
	Figure~\ref{fig:Cf}(b) shows the vertical downward shift in the logarithmic region for all test surfaces at a matched $Re_{\tau} \approx 27700$ (the highest matched $Re_{\tau}$ for \textbf{PR2} set and second highest for \textbf{PR1}) from a smooth wall TBL at this matched $Re_{\tau}$. This shift, denoted by $\Updelta U^+$, is shown as a function of $k_s^+$ in figure~\ref{fig:Cf}(c) for all test surfaces (white-filled contours). All collapse to the logarithmic function $1/\kappa \ln k_s^+ + B - B_{FR}$ (equation~\ref{eq:ks}), although this is given since $k_s$ is not obtained independently from $\Delta U^+$. The magnitude of $k_s^+$ (table~\ref{tab:case}) exceeds the threshold defined by \cite{flack2010}, $k_s^+ \gtrsim 70$, ensuring that all test surfaces are `fully' rough and thus $\Updelta U^+$ depends solely on $k_s$. It should be noted that $\Delta U^+$ here accounts for the total momentum deficit (from that of smooth wall TBLs) in both porous and rough walls, and it shows that it is possible to characterise porous wall with the same framework used for rough wall characterisation (i.e. logarithmic shift that scales on $k_s$). The effect of permeability on $\Updelta U^+$ is shown in figure~\ref{fig:Cf}(c) for test surfaces comprises of porous walls (\textbf{P}, \textbf{PR1}, and \textbf{PR2}, filled contours). Although it is clear that $\Updelta U^+$ is logarithmically scaled by $Re_K$, universality, as shown in figure~\ref{fig:Cf}(c) with $k_s$, is not apparent. This highlights the importance of accounting for the `blockage' effect (equation~\ref{eq:porous4}). At high range of $Re_{\tau}$ tested in this study, the outer layer similarity is preserved for all test surfaces (i.e. velocity defect profiles collapse beyond $y/\delta \gtrsim 0.2$, figure \ref{fig:Cf}d), ensuring that a given value of $k_s$ (for a surface) can be used to predict the drag at higher Reynolds numbers. 
	
	%%%%%%%%%%%%%%%%%%%%%%%%%%%%%%%%%%%%%%%%%%%%%%%%%%%%%%%%%%%%%%%%%%%%%%%%%%%%%%%%%%%%%%%%%%%%%%%%
	\section{Decoupling permeability and roughness effect}
	\label{sub:decouple}
	
	\subsection{Additional blockage effect}
	\label{sub:method1}
	We rewrite equation (\ref{eq:ksp}) for porous wall (case \textbf{P}) in a more general term
	\begin{equation}
		k_{s_p} = Re_K f(k_{s_b})
		\label{eq:ksp_gen}
	\end{equation}
	where $f(k_{s_b})$ is a function representing the blockage effect by the substrate. As a roughness interface is added (case \textbf{PR}), a roughness term, possibly a function of the characteristic lengthscale of roughness (i.e. rough wall $k_s$) increases blockage 
	\begin{equation}
		k_{s_{pr}} = Re_K f(k_{s_b}, k_{s_r})
		\label{eq:kspr_gen}
	\end{equation}
	where subscripts `$r$' denotes the rough wall and `$pr$' the porous-rough wall. When two substrates have an approximately matched permeability, then the additional blockage effect can be approximated by $f(k_{s_b},k_{s_r})/f(k_{s_b}) \approx k_{s_{pr}}/k_{s_p}$. Case \textbf{PR1} has 5 matched $Re_K$ with case \textbf{P} (within 5\% difference, table \ref{tab:case}). As the $k_s$ for these surfaces are known from fitting $\Updelta U^+$ (\S\ref{sub:wss}), the additional blockage is approximated by $f(k_{s_b},k_{s_r})/f(k_{s_b}) = 2.25$ (figure \ref{fig:Cf}e) within 10\% error due to $Re_K$ difference. This is interpreted as the rough wall \textbf{R1} increases the total blockage effect by approximately 2.25 times from that of porous wall.
	
	At this point, the empirical formulation of $f(k_{s_b},k_{s_r})$ is unknown. Let the roughness effect be represented solely by rough wall $k_s$
	\begin{equation}
		f(k_{s_b},k_{s_r}) = f(k_{s_b}) C k_{s_r}
	\end{equation}
	where $C$ is a constant. For an approximately matched $Re_K$, $C \approx k_{s_{pr}}/(k_{sp} k_{s_r})$. We obtain $C = 0.17$ for case \textbf{PR1} (figure \ref{fig:Cf}e), and further test this formulation for the second roughness \textbf{PR2}. Here, there are 4 matched $Re_K$ cases (within 10\% difference, table \ref{tab:case}). The total blockage effect increases by approximately 2.5 times from the porous wall (figure \ref{fig:Cf}e), consistent with the increasing $C_f$ of case \textbf{R2} from that of \textbf{R1} (figure \ref{fig:Cf}a). The constant $C$ appears to be similar to that obtained from \textbf{R1}, $C = 0.15$. This suggest that the formulation has the potential to be applicable across different cases and is not dependent on the type of roughness. 
	
	\subsection{Equivalent homogeneous roughness}
	\label{sub:method2}
	
	A recent study by \cite{hutchins2023} suggested that for a heterogeneous rough wall (i.e. a rough wall constructed from various patches of homogeneous roughness of roughness length scale $k_{s_i}$ covering an area $A_i$), the equivalent homogeneous roughness length $k_{ehr}$ can be defined as:
	\begin{equation}
		k_{ehr} = \left[ \frac{1}{A} \sum_{i = 1}^{N} k^n_{s_i} A_i \right]^{\frac{1}{n}} \quad \mathrm{where} \quad A = \sum_{i = 1}^{N} A_i
		\label{eq:kehr}
	\end{equation}
	In the present study, the drag penalty from all test surfaces can be represented by the logarithmic shift $\Updelta U^+$ (figure \ref{fig:Cf}b), suggesting the possibility of characterising permeable walls with the same framework used for rough wall TBLs. Thus, we might consider porous-rough walls as simply a combination of two `homogeneous roughnesses' overlaying each other. Equation (\ref{eq:kehr}) is therefore reduced to
	\begin{equation}
		k_{s_{pr}} = \left[ k_{s_p}^n + k_{s_r}^n\right]^{\frac{1}{n}}
		\label{eq:kspr}
	\end{equation}
	For case \textbf{PR1}, $n$ is obtained by solving equation (\ref{eq:kspr}) with the known $k_s$ for each \textbf{PR1}, \textbf{P}, and \textbf{R1}, resulting in $n = 1.42$. The same approach is applied to case \textbf{PR2}, resulting in $n = 1.54$, which is relatively consistent with that obtained from case \textbf{PR1}. %This shows that the effect of roughness  is stratified even when the porous substrate thickness is much greater than the pore size ($k/s \approx 17$, figure \ref{fig:setup}d).
	
	It is still unknown, at this point, what the empirical formulation for the roughness effect is. Present results suggest that it is possibly in terms of an additional logarithmic shift similar to equation (\ref{eq:porous4}). Further, it is still unclear what the empirical relation with the characteristic lengthscale of roughness (rough wall $k_s$) is. For the two methods tested in \S\ref{sub:decouple}, it is neither understood what the physical interpretations of constants $C$ and $n$ are, nor whether these constants are universal for various types of rough walls. We are unable to answer these questions in this study. However, present results suggest a promising start towards decoupling permeability and roughness effects.
	
	%\begin{table}
	%    \centering
	%    \begin{tabular}{c c c}
	%         Parameter &  \textbf{PR1} &\textbf{PR2}\\
	%         \hline
	%         $k_{s_{pr}}/k_{s_p}$ & 2.3 & 2.5\\
	%         $C$ & 0.16 & 0.15\\
	%         $n$ & 1.42 & 1.53
	%    \end{tabular}
	%    \caption{Decoupling of permeability and roughness effect using methods described in \S\ref{sub:decouple}.}
	%    \label{tab:fit}
	%\end{table}
	
	%%%%%%%%%%%%%%%%%%%%%%%%%%%%%%%%%%%%%%%%%%%%%%%%%%%%%%%%%%%%%%%%%%%%%%%%%%%%%%%%%%%%%%%%%%%%%%%%
	\section{Conclusions and future work}
	
	We conduct velocity and drag measurements of TBLs developing over various porous-rough wall combinations, with the roughness effect systematically varied while maintaining the permeability effect. Measurements are conducted at relatively high Reynolds numbers($14400 \leq Re_{\tau} \leq 33100$), with a set of matched $Re_{\tau}$ (within 10\%) and $Re_K$ ($13 \leq Re_K \leq 27.5$) of each porous-rough combinations. Present results suggest that the increase in drag over porous, rough, and porous-rough walls is characterised by a downward shift in the logarithmic region $\Updelta U^+$ from that of smooth wall TBLs and that the mean flow follows outer-layer similarity for cases examined at these high Reynolds numbers (with substantial scale separation). Further analysis shows that $\Updelta U^+$ from these surfaces can be characterised by the roughness lengthscale $k_s$, suggesting the possibility of characterising porous walls with the same  framework used for rough wall TBLs. Following the hypothesis of \cite{esteban2022}, it is viable that for a porous-rough wall, permeability and roughness effects are represented by $Re_K$ and a total blockage function $f(k_{s_b},k_{s_r})$, respectively, with the total blockage consists of the blockage effect from the porous substrate and additional roughness interface above the porous wall. Present results suggest that the rough walls increase the blockage of porous-rough walls by approximately 2.3--2.5 times from that of the porous wall for both roughnesses tested in this study. We observed that the additional roughness effect follows $C k_{s_r}$, where constant $C \approx 0.16$ for both rough wall test surfaces. Analysis of the same data with the method suggested by \cite{hutchins2023} shows that the effect of porous and rough walls may be decoupled using an area-weighted power-mean (with $n \approx$ 1.5 for both surfaces) to obtain an equivalent roughness length for both rough wall test surfaces. The empirical formulation as well as the physical interpretation of constants observed in this study requires further data and should be the focus of future work. 
	%%%%%%%%%%%%%%%%%%%%%%%%%%%%%%%%%%%%%%%%%%%%%%%%%%%%%%%%%%%%%%%%%%%%%%%%%%
	
	\backsection[Funding]{We gratefully acknowledge the financial support from EPSRC (Grant Ref no: EP/S013296/1) and European Office for Airforce Research and Development (Grant No: FA9550-19-1-7022, Programme Manager: Dr. Doug Smith).}
	
	\backsection[Declaration of interest]{The authors declare no conflict of interest.}
	
	\backsection[Data availability statement]{The data that support the findings of this study will be made available upon publication.}
	
	\backsection[Author ORCID]{D. D. Wangsawijaya: \href{https://orcid.org/0000-0002-7072-4245}{0000-0002-7072-4245}, P. Jaiswal: \href{https://orcid.org/0000-0002-5240-9911}{0000-0002-5240-9911}, B. Ganapathisubramani: \href{https://orcid.org/0000-0001-9817-0486}{0000-0001-9817-0486}}
	
	\backsection[Author contributions]{PJ and DDW designed, carried out measurements, and post-processed the data. DDW analysed the data and wrote the manuscript, BG was responsible for conceptualisation, funding acquisition, editing of drafts and project management.}
	
	%%%%%%%%%%%%%%%%%%%%%%%%%%%%%%%%%%%%%%%%%%%%%%%%%%%%%%%%%%%%%%%%%%%%%%%%%%%%%%%%%%%%%%%%%%%%%%%%
	%\bibliographystyle{jfm}
	%\bibliography{jfm}
	%Use of the above commands will create a bibliography using the .bib file. Shown below is a bibliography built from individual items.
	
	\bibliographystyle{jfm}
	\bibliography{biblio}

	%% End of file `jfm2esam.bib'.
	
\end{document}